\documentclass[12pt,a4paper]{article}
\usepackage{mathtools}
\usepackage{mathrsfs}
\usepackage{amsmath}
\usepackage{amsfonts}
\usepackage{tikz-feynman}
\usepackage[utf8x]{inputenc}
\usepackage{amssymb}
\usepackage{slashed}
\usepackage{bbm}
\usepackage{bm}
\usepackage{color}
\usepackage[a4paper,top=3cm,bottom=3cm,left=2cm,right=3cm,bindingoffset=5mm]{geometry}
\usepackage{booktabs} 
\usepackage{tikz}
\usepackage{cancel}
\usepackage[bottom]{footmisc}
\usepackage{cite}
\usepackage{float}
\usepackage{todonotes}
\usepackage{jheppub}
\usepackage{bbold}

\usetikzlibrary{shapes.misc}
\tikzset{
    point/.style={
    draw=black,
    cross out,
    inner sep=0pt,
    minimum width=4pt,
    minimum height=4pt,
    },
}

\usepackage{hyperref}

\hypersetup{
colorlinks=true,         
linkcolor=blue,          
citecolor=red,        
urlcolor=cyan            
}

\renewcommand{\[}{\begin{equation}\begin{aligned}}
\renewcommand{\]}{\end{aligned}\end{equation}}

\newcommand{\wb}{{\bar{w}}}

\title{
No U(1) `electric-magnetic' duality in Einstein gravity
}
\author{Ricardo Monteiro}

\affiliation{Centre for Theoretical Physics, Department of Physics and Astronomy, \\ Queen Mary University of London, E1 4NS, United Kingdom}

\emailAdd{ricardo.monteiro@qmul.ac.uk}

\abstract{
We revisit the question of whether classical general relativity obeys, beyond the linearised order, an analogue of the global U(1) electric-magnetic duality of Maxwell theory, with the Riemann tensor playing the role analogous to the field strength. Following contradictory claims in the literature, we present a simple gauge-invariant argument that the duality does not hold. The duality condition is the conservation of the helicity charge. Scattering amplitudes of gravitons in general relativity, and of gluons in Yang-Mills theory, violate this selection rule already at tree level. Indeed, the maximally-helicity-violating (MHV) amplitudes are famous for their simplicity. The duality in the linearised theories is, therefore, broken by the interactions. In contrast, the tree-level scattering amplitudes in duality-invariant theories of non-linear electromagnetism are known to obey helicity conservation. While the duality is not a symmetry of the full theory of general relativity, it does hold within a sector of the solution space, including vacuum type D solutions, where the duality is known to rotate between mass and NUT charge.
}

\begin{document}

\begin{flushright}
QMUL-PH-23-24
\end{flushright}

\maketitle


\section{Introduction}

The electric-magnetic duality of electromagnetism has long been a source of inspiration for theoretical developments. Here, we are interested in investigating the extension of the U(1) duality to other theories.\footnote{We will not discuss the related weak coupling / strong coupling duality, which is famous in the context of supersymmetric non-Abelian theories \cite{Montonen:1977sn,Seiberg:1994pq}.} The duality is known to extend to a class of theories of non-linear electromagnetism identified in \cite{Gaillard:1981rj}. In general relativity, a version of the duality where the Riemann tensor plays the role analogous to the field strength is known to hold at linearised level; see e.g.~\cite{Penrose:1960eq,Nieto:1999pn,Hull:2001iu,Henneaux:2004jw,Deser:2005sz,Bunster:2006rt,Argurio:2009xr,Barnich:2008ts,deHaro:2008gp,Astorino:2019ljy,Boos:2021suz,Astorino:2023elf}. However, ref.~\cite{Deser:2005sz} reported that the first non-linear correction, i.e.~the cubic vertex, does not admit a duality-invariant formulation, similarly to the case of Yang-Mills theory \cite{Deser:1976iy}. The procedure employed was not covariant, and we will see that, at least in our approach, the cubic vertex is subtle regarding the duality. It has since been suggested in various instances, in writing or not, that perhaps a non-linear realisation of the gravitational duality is possible. In fact, there is a recent claim that this is the case \cite{Kol:2022bsd,Kol:2023yxd}, partly motivated by --- but not implied by --- developments in scattering amplitudes \cite{Luna:2015paa,Huang:2019cja,Alawadhi:2019urr,Banerjee:2019saj,Emond:2020lwi,Moynihan:2020gxj,Monteiro:2020plf,Crawley:2021auj,Monteiro:2021ztt}. See also \cite{Compere:2022zdz} for another non-linear construction that was proposed to represent the duality.

In this paper, we show that the duality does not hold in the full theory of general relativity, but explain also why it holds within a certain sector of solutions to the vacuum Einstein equations, which includes type D vacuum solutions. Indeed, it is well known that the Kerr-Taub-NUT family of solutions admits the U(1) duality, rotating between mass and NUT charge. This is an analogue of the rotation between electric charge and magnetic monopole charge in electromagnetism.

The understanding of the duality that we employ builds on work that goes a long way back, to the realisation that the duality's Noether symmetry is helicity conservation \cite{Calkin:1965}. This can be seen as the invariance of an action for the positive and negative helicity degrees of freedom of the gauge boson under chiral U(1) transformations. Helicity conservation is expressed in a gauge-invariant manner as a selection rule on the scattering amplitudes: the only admissible non-trivial amplitudes have an equal number of positive and negative helicity external particles (in a convention where all particles are incoming, say). The scattering amplitudes of gravitons in Einstein's theory, or of gluons in Yang-Mills theory, famously violate this rule. Indeed, the simplest and best known tree-level amplitudes in these theories are classified as MHV --- maximally-helicity-violating amplitudes. On the other hand, it was shown in \cite{Novotny:2018iph} that all duality-invariant theories of non-linear electrodynamics satisfy the selection rule, generalising an earlier observation in Born-Infeld theory \cite{Rosly:2002jt}. Analogous results have been obtained in extended supergravities, for the duality acting on a class of spin-1 particles in the theory (and a quantum anomaly occurs which impacts the ultraviolet behaviour), as well as in related contexts \cite{Carrasco:2013ypa,Bern:2013uka,Freedman:2017zgq,Bern:2017rjw,Pavao:2022kog,Carrasco:2022jxn,Carrasco:2023qgz}. Our discussion includes a straightforward explanation of this correspondence between the U(1) duality and the amplitudes selection rule.

The paper is organised as follows. In section~\ref{sec:maxwell}, we review the original duality in Maxwell theory, and the associated conservation of helicity. In section~\ref{sec:nonlinEM}, we discuss the extension of the duality to non-linear electromagnetism, and its manifestation in terms of a selection rule on the scattering amplitudes. In section~\ref{sec:argument}, we present the general argument for the selection rule as the condition for the duality to hold. In sections~\ref{sec:YM} and \ref{sec:grav}, we describe the failure of the duality in Yang-Mills theory and in general relativity, respectively. In section~\ref{sec:typeD}, we discuss a class of algebraically special gravity solutions where the duality holds, manifesting itself as a symmetry that rotates between mass and NUT charge. We conclude with brief remarks in section~\ref{sec:conclusion}.


\section{Maxwell theory}
\label{sec:maxwell}

The U(1) electric-magnetic duality of Maxwell theory is the statement that the transformation
\begin{align}
    \begin{pmatrix}
           F_{\mu\nu} \\
           \ast F_{\mu\nu}
            \end{pmatrix} \mapsto \begin{pmatrix}
           F'_{\mu\nu} \\
           \ast F'_{\mu\nu}
            \end{pmatrix} =
         \begin{pmatrix}
           \cos\theta & \sin\theta \\
          -\sin\theta & \cos\theta
            \end{pmatrix} \begin{pmatrix}
           F_{\mu\nu} \\
           \ast F_{\mu\nu}
            \end{pmatrix}
  \end{align} 
preserves the Maxwell equations,
\[
d \!\ast \!F=0\,, \quad dF=0\,.
\]
Here, $\ast$ denotes the Hodge dual, such that $ \ast F_{\mu\nu}=\frac1{2}\varepsilon_{\mu\nu\alpha\beta}F^{\alpha\beta}$. In Lorentzian signature, $\ast^2=-1$.
The chiral nature of the transformation is clear when we express it in terms of the (anti-)self-dual parts of the field strength,\footnote{This convention will mean that, in later sections, we will see $F_+\sim e-ig$ for the dyon field strength, and $\psi\sim m-in$ for the Taub-NUT Weyl spinor.\label{footnote}}
\[
F_\pm=\frac1{2} (F \mp i \ast F) \quad \Leftrightarrow \quad
\ast F_\pm = \pm i F_\pm\,,
\]
for which the transformation is
\[
\label{eq:U1F}
F_\pm \mapsto e^{\pm i\theta} F_\pm.
\]
In our convention, a self-dual field has positive helicity, and an anti-self-dual field has negative helicity.

This U(1) transformation is a symmetry of the equations of motion, though not of the usual action:
\[
S[A_\mu] = -\frac1{4} \int d^4x\, F^{\mu\nu}F_{\mu\nu} \; \mapsto \;  \cos(2\theta) \,S[A_\mu]\,,
\]
where we dropped a boundary term. We can easily write an action that makes the duality manifest in terms of the chiral transformation \eqref{eq:U1F}, by using light-cone gauge. Let us consider double-null coordinates, such that
\[
ds^2=2(-du dv+dw d\wb)\,, \quad \square=2(-\partial_u\partial_v+\partial_w\partial_\wb)\,.
\]
Choosing the light-cone gauge condition $A_u=0$, we obtain the action
\[
S[A_v,A_w,A_\wb] =   \int d^4x\, \left(\,
\frac1{2} \,(\partial_u A_v-\partial_\wb A_w-\partial_w A_\wb)^2 + A_w \square A_\wb
\right)\,.
\]
The component $A_v$ can be integrated out exactly, leading to an action for the two physical degrees of freedom. These are identified with the positive and negative helicity fields, which in our convention are
\[
A_+=A_\wb\,, \quad A_- = A_w\,.
\]
Hence, we obtain the action
\[
\label{eq:actionEM}
S[A_\pm] =   \int d^4x\, A_-\, \square\, A_+ \,,
\]
which is clearly invariant under the chiral transformation
\[
\label{eq:U1A}
A_\pm \mapsto e^{\pm i\theta} A_\pm\,.
\]
The associated Noether current is
\[
j^\mu = i(A_+\partial^\mu A_- - A_-\partial^\mu A_+)\,,
\]
and the conserved charge is
\[
{\mathcal H} = i \int d^3 x\, (A_+\partial_t A_- - A_-\partial_t A_+)\,.
\]
To interpret this chiral symmetry, we can use the second quantisation formalism, where we take $A_\pm$ to be operators built with creation and annihilation operators in Fourier space. This leads after normal ordering to
 \[
\mathcal H = \int \frac{d^3 k}{(2\pi)^3\, 2|\vec k|}\, (a^\dagger_{+,k}\,a_{+,k}-a^\dagger_{-,k}\,a_{-,k})
= N_+- N_-\,.
\]
Here, $a_{\pm,k}$ annihilate a photon of helicity $\pm$ and momentum $k_\mu$, 
while $ N_\pm$ are the particle number operators for the given helicity. So the U(1) symmetry implies the conservation of the total helicity. To our knowledge, this observation was first made in a clear manner in ref.~\cite{Calkin:1965}. The story is analogous to that of the chiral symmetry of massless fermions. In fact, the analogy extends to quantum effects: ref.~\cite{Agullo:2016lkj} has shown that the symmetry is anomalous on a curved background, such that $\langle\nabla_\mu j^\mu\rangle \sim R_{\mu\nu\alpha\beta}\ast\! \!R^{\mu\nu\alpha\beta}$, similarly to the fermionic chiral anomaly.


\section{Non-linear electrodynamics}
\label{sec:nonlinEM}

Let us consider now non-linear extensions of electromagnetism, where the Lagrangian density is a function of the field strength,
\[
\label{eq:LF}
{\mathcal L}(F) = -\frac1{4} \, F^{\mu\nu}F_{\mu\nu} + {\mathcal O}(F^4)\,.
\]
If the interactions in ${\mathcal O}(F^4)$ are to preserve the symmetry under $A_\pm \mapsto e^{\pm i\theta} A_\pm$, they must take the following schematic form in the action for the two degrees of freedom:
\[
\label{eq:actionnonlinEM}
S[A_\pm] =   \int d^4x \,\Big( \,A_-\, \square\, A_+ + \sum_{n\geq2}\, \frac1{M^{4(n-1)}}\, \partial^{2n} A_-^n A_+^n \,\Big) \,,
\]
where $M$ has units of mass.\footnote{When we schematically write $\partial^{2n}$ in \eqref{eq:actionnonlinEM}, this includes the possibility of inverse derivatives, with $2n$ being the net derivative order. As we will see later in the well-known cases of Yang-Mills and gravity, the integral operation $1/\partial_u$  is typical of interactions in light-cone gauge actions, leading to the appearance of  $\partial^{2n+m}/\partial^m_u$. We recall that $u$ is the light-cone coordinate associated to the gauge choice. This feature presents no difficulties when considering perturbation theory in Fourier space. Regarding Lorentz symmetry, there is a well-established machinery to test this property for an action of the type \eqref{eq:actionnonlinEM}, going back to ref.~\cite{Bengtsson:1983pg}.} In the Feynman diagrams obtained from such an action, helicity is manifestly conserved at each vertex. Putting the action into this form, however, may require non-trivial field redefinitions on top of the gauge choice. The invariant statement is that the scattering amplitudes (with all external particles taken to be, say, incoming) must obey the selection rule
\[
\label{eq:Aduality}
\boxed{ \;{\mathcal A}(n_+,n_-)=0\quad \text{for} \quad n_+\neq n_-\; }\,,
\]
where $n_\pm$ are the number of $\pm$ helicity external particles. That is, the amplitudes with unequal number of positive and negative helicity external particles must vanish. Conversely, if the amplitudes satisfy this selection rule, then using the explicit form of the non-vanishing amplitudes we can in principle construct an action of the form \eqref{eq:actionnonlinEM}.

Theories of non-linear electromagnetism that admit the electric-magnetic duality are identified by the Noether-Gaillard-Zumino condition \cite{Gaillard:1981rj,Gibbons:1995cv}:
\[
G_{\mu\nu} \ast \!G^{\mu\nu}
- F_{\mu\nu} \ast \!F^{\mu\nu} = C\,,
\quad \text{with} \quad G_{\mu\nu} = -2 \frac{\partial {\mathcal L}}{\partial F^{\mu\nu}}\,,
\]
and where $C$ is a constant. This constant vanishes if the action is of the type \eqref{eq:LF}, i.e.~if $G$ is identified with $F$ at leading order. The duality acts as
\begin{align}
    \begin{pmatrix}
           G_{\mu\nu} \\
           \ast F_{\mu\nu}
            \end{pmatrix} \mapsto \begin{pmatrix}
           G'_{\mu\nu} \\
           \ast F'_{\mu\nu}
            \end{pmatrix} =
         \begin{pmatrix}
           \cos\theta & \sin\theta \\
          -\sin\theta & \cos\theta
            \end{pmatrix} \begin{pmatrix}
           G_{\mu\nu} \\
           \ast F_{\mu\nu}
            \end{pmatrix}\,,         
  \end{align} 
  preserving the equation of motion, $d \!\ast \!G=0$\,, and the Bianchi identity, $dF=0$\,. Ref.~\cite{Novotny:2018iph} has shown that the theories of non-linear electromagnetism admitting the electric-magnetic duality (with $C=0$) satisfy helicity conservation at tree level, that is, obey the selection rule \eqref{eq:Aduality}.\footnote{The Born-Infeld theory is an example of a non-linear electromagnetism model satisfying the duality condition and, indeed, its tree-level amplitudes satisfy the selection rule, which was first determined in \cite{Rosly:2002jt}. Using unitarity methods, ref.~\cite{Elvang:2019twd} reported the existence of helicity-violating amplitudes at one loop; see also \cite{Wen:2020qrj,Elvang:2020kuj}. From the point of view of the action \eqref{eq:actionnonlinEM}, this loop-level U(1) violation presumably arises from the fact that the field redefinitions required to put the action into this form affect the measure of the path integral.}


\section{The general argument}
\label{sec:argument}

The empirical fact that scattering amplitudes in duality-satisfying theories of non-linear electrodynamics obey the selection rule \eqref{eq:Aduality} admits a simple explanation. In fact, the selection rule is the condition for `electric-magnetic' duality in a broader range of theories with massless particles.

An obvious difficulty in theories like Yang-Mills or gravity is that, beyond the linearised level (where the duality is straightforward, as we will review shortly), it is unclear how to implement the U(1) transformation. This is why the perspective of scattering amplitudes is helpful: if it is clear how to implement the transformation at linearised level, then it is clear how it acts on the asymptotic scattering states, i.e.~the initial and final particles in the amplitude. The U(1) transformation with parameter $\theta$ is implemented via a unitary operator, $e^{i\theta{\mathcal H}}$; we saw that, for photons, this operator is $e^{i\theta(N_+-N_-)}$. Now, the S-matrix contains all the information about the time evolution. While we do not know the S-matrix of quantum gravity, we know in principle all the information required for classical gravity. The duality is equivalent to the invariance of S-matrix elements under the duality:
\[
\label{eq:SH}
\langle \text{out}_\theta | S | \text{in}_\theta\rangle =  \langle \text{out} |e^{-i\theta{\mathcal H}} S e^{i\theta{\mathcal H}}| \text{in}\rangle =\langle \text{out} | S | \text{in}\rangle \,.
\]
That is, the operator ${\mathcal H}$ commutes with the scattering operator $S$. 

For now, let us consider only one massless particle species, say the scattering of photons in non-linear electromagnetism, of gluons in Yang-Mills theory, or of gravitons in general relativity. For these, the U(1) operator is based on the helicity charge, $e^{i\theta{\mathcal H}} = e^{i\theta(N_+-N_-)}$.\footnote{Since ${\mathcal H}$ represents the helicity charge, one may instead wish to normalise it such that $e^{i \theta{\mathcal H}}=e^{i \theta s(N_+-N_-)}$, where $s$ is the spin of the particle, e.g.~$s=2$ for gravitons. We will not do this, as it won't affect the subsequent discussion. One may absorb $s$ into the duality parameter $\theta$.} Clearly, it is easier to look at amplitudes for initial and final particles that have definite positive or negative helicity. Now, in the scattering amplitudes literature, making use of crossing symmetry, external particles are often taken as being all incoming (or outgoing).\footnote{A massless particle that has helicity $\pm$ seen as incoming, has helicity $\mp$ seen as outgoing.} The amplitude is then defined as
\[
{\mathcal A} = \langle 0 | T | \text{in} \rangle \,, \quad \text{with} \quad S=1+iT\,.
\]
Under the U(1) transformation, we have
\begin{align}
{\mathcal A}_\theta = \langle 0 | T |\text{in}_\theta\rangle  = \langle 0 | T e^{i \theta(N_+-N_-)}|\text{in}\rangle  = e^{i \theta(n_+-n_-)} \langle 0 | T |\text{in}\rangle= e^{i \theta(n_+-n_-)} {\mathcal A}\,,
\label{eq:ampruleder}
\end{align}
where $n_\pm$ are the number of $\pm$ helicity particles in the asymptotic state $|\text{in}\rangle$.
The duality is the statement that the amplitudes are invariant under U(1). This is, therefore, equivalent to the selection rule \eqref{eq:Aduality} seen previously.

The argument applies more generally. For instance, in Einstein-Maxwell theory, the electric-magnetic duality holds (classically) as a U(1) symmetry acting on the gauge field. The selection rule \eqref{eq:Aduality} then applies to tree amplitudes for photons in this theory, which comes out naturally from the formulas in \cite{Cachazo:2014xea}; see also \cite{Bern:1999bx,Chiodaroli:2014xia} for related work.\footnote{The amplitudes studied in these works, obtained as a `double copy', actually correspond to Einstein-dilaton-axion-Maxwell theory, which obeys an SL(2,$\mathbb{R}$) extension of the electric-magnetic duality \cite{Gaillard:1981rj,Gibbons:1995ap}.} This should extend to duality-preserving higher-derivative extensions of Einstein-Maxwell theory, as those studied in \cite{Cano:2021tfs}.

Another natural setting is to consider the coupling to massive particles, and the duality is still the statement that the amplitudes are U(1) invariant. In the Maxwell case, we can introduce dyonic particles, with electric charge $e$ and magnetic monopole charge $g$. The gauge field generated by a static dyon is
\[
\label{eq:Fdyon}
F=dA=\frac{e}{r^2}\, dt\wedge dr -   \frac{g}{r^2}\, \ast (dt\wedge dr)\,.
\]
By considering the action of U(1) on $F$, we can determine the U(1) transformation of the charges:
\[
F_+ \,\mapsto\, e^{i\theta} F_+ \quad \Rightarrow \quad e- i g \,\mapsto\, e^{ i\theta}(e- i g)= e'- i g'\,.
\]
Now, the 3-point amplitude for a dyon to absorb a $\pm$ helicity photon is \cite{Huang:2019cja}
\[
\label{eq:A3EM}
{\mathcal A}= (e\pm ig)\; \varepsilon_\pm\cdot p\,,
\]
where $p$ is the momentum of the dyon (say before the interaction), and $\varepsilon_\pm$ is the polarisation vector of the $\pm$ helicity incoming photon.\footnote{While 3-point on-shell kinematics requires complexified momenta in Lorentzian signature, 3-point amplitudes have proven extremely useful, e.g.~for constructing higher-point or even higher-loop amplitudes via recursion \cite{Britto:2005fq,Arkani-Hamed:2012zlh} and unitarity methods \cite{Bern:2011qt}. In the case of dyons, higher-point amplitudes are poorly understood due to the well-known difficulties with magnetic monopoles; see \cite{Csaki:2020inw} for some recent progress. Regarding \eqref{eq:A3EM} and \eqref{eq:Fdyon}, it was shown in \cite{Monteiro:2020plf,Crawley:2021auj,Monteiro:2021ztt}, based on the KMOC formalism \cite{Kosower:2018adc}, how they are related via an on-shell Fourier transform.
}
 Notice the dyon couples differently to the two helicities of the photon. This amplitude is invariant under U(1), because the phases cancel between the coupling and the photon. In gravity, there is an analogue of the electromagnetic dyon, as we will review later, where the mass and the NUT charge play roles analogous to the electric and magnetic monopole charges. This will not imply, however, that the full interacting theory is duality invariant. In fact, neither the 3-point amplitude \eqref{eq:A3EM} nor its gravitational analogue probe interactions between the massless gauge bosons, which are absent in the case of electromagnetism.

To conclude this section, notice that if we demand only the ${\mathbb Z}_2$ electric-magnetic duality, $F\mapsto \ast F$, which corresponds to \,$\theta=\pi/2$\,, then the selection rule \eqref{eq:Aduality} is too strong. From  \eqref{eq:ampruleder}, we see that the requirement $n_+=n_-$ is substituted by $(n_+-n_-)/4\in {\mathbb Z}$. This weaker requirement will not alter the failure of the duality in Yang-Mills theory and in gravity, which we will discuss in the following sections.


\section{Yang-Mills theory}
\label{sec:YM}

At linearised level, it is clear that Yang-Mills theory obeys the duality, because Maxwell theory does. At non-linear level, our argument implies that the duality fails, because the scattering amplitudes do not obey the selection rule of helicity conservation. In fact, there is a famous class of tree-level amplitudes presenting maximal helicity violation (MHV), where all external gluons have positive helicity except for two. It admits a very simple formula at $n$ points \cite{Parke:1986gb}: in spinor-helicity language,
\[
\label{eq:PT}
{\mathcal A}(i^-,j^-,\text{rest}^+) =  g^{n-2}\, \langle ij\rangle^4 \left( \frac{\text{Tr}(T^{a_1}T^{a_2}T^{a_3}\cdots T^{a_n})}{\langle 12\rangle \langle 23\rangle \cdots \langle n1\rangle} + \text{perm} \right)\,,
\]
where `perm' denotes a sum over non-cyclic permutations of the $n$ external legs. Obviously, there is a conjugate $\overline{\text{MHV}}$ class of amplitudes, as the theory is parity invariant. Tree amplitudes with all or (except for 3 points) all-but-one particles of the same helicity vanish, hence the name MHV for the all-but-two class above.

If we write down the light-cone gauge action \cite{Chalmers:1998jb}
\begin{align}
S[A_\pm] = \int d^4x \,  \text{Tr}\Bigg(  & A_- \square A_ + + 2g\Big(\frac{\partial_w}{\partial_u}A_+\Big) i [\partial_u A_-,A_+]
+2g\Big(\frac{\partial_\wb}{\partial_u}A_-\Big) i [\partial_u A_+,A_-] \nonumber \\
& - 2g^2[\partial_u A_-,A_+] \frac1{\partial_u^2} [\partial_u A_+,A_-]
 \Bigg)\,,
\end{align}
we see that it is not invariant under $A_\pm \mapsto e^{\pm i\theta} A_\pm$.\footnote{As already mentioned, the apparent non-locality here is a feature of light-cone gauge. This, on top of the lack of manifest Lorentz invariance, is the price to pay for reducing to two physical off-shell degrees of freedom.} Field redefinitions cannot help, due to the invariant statement that amplitudes such as \eqref{eq:PT} violate the duality selection rule. In fact, the existence of non-vanishing amplitudes for an odd number of particles already invalidates the duality.


\section{Gravity}
\label{sec:grav}

The `electric-magnetic' duality in linearised gravity works similarly to the case of Maxwell theory; see e.g.~\cite{Henneaux:2004jw}. The Riemann tensor plays the role analogous to the field strength. The linearised Riemann tensor is
\[
R_{\mu \nu\lambda\rho} =  \frac1{2}\left(  \partial_\mu\partial_\rho h_{\lambda\nu}
- \partial_\nu\partial_\rho h_{\lambda\mu}
- \partial_\mu\partial_\lambda h_{\rho\nu} 
+ \partial_\nu\partial_\lambda h_{\rho\mu}   \right),
\label{eq:Rlin}
\]
and its Hodge dual is
\[
\ast R_{\mu \nu\lambda\rho } = \frac1{2}\, \varepsilon_{\mu \nu\alpha\beta} \,R^{\alpha\beta}{}_{\lambda\rho }\,.
\]
The equation of motion is the vacuum Einstein equation,
\[
R^\lambda{}_{\mu\lambda\nu}=0\,,
\]
while the first (algebraic) and second (differential) Bianchi identities are, respectively,
\[
R_{\mu [\nu\lambda\rho]}=0\,, \qquad \partial_{[\mu}R_{\nu \lambda]\rho\alpha}=0\,.
\]
The duality is the invariance of the equation of motion and the Bianchi identities under
\begin{align}
    \begin{pmatrix}
           R_{\mu \nu\lambda\rho } \\
           \ast R_{\mu \nu\lambda\rho }
            \end{pmatrix} \mapsto \begin{pmatrix}
           R'_{\mu \nu\lambda\rho } \\
           \ast R'_{\mu \nu\lambda\rho }
            \end{pmatrix} =
         \begin{pmatrix}
           \cos\theta & \sin\theta \\
          -\sin\theta & \cos\theta
            \end{pmatrix} \begin{pmatrix}
           R_{\mu \nu\lambda\rho } \\
           \ast R_{\mu \nu\lambda\rho }
            \end{pmatrix}\,.
            \label{eq:dualityR}
  \end{align} 
In terms of the (anti-)self-dual parts of the curvature,
\[
\label{eq:asdR}
R^\pm_{\mu \nu\lambda\rho }=\frac1{2} (R_{\mu \nu\lambda\rho } \mp i \ast \! R_{\mu \nu\lambda\rho }) \quad \Leftrightarrow \quad
\ast R^\pm_{\mu \nu\lambda\rho } = \pm i R^\pm_{\mu \nu\lambda\rho }\,,
\]
we have the expected chiral transformation,
\[
\label{eq:U1R}
R^\pm_{\mu \nu\lambda\rho } \mapsto e^{\pm i\theta} R^\pm_{\mu \nu\lambda\rho}\,.
\]

For the non-linear problem, let us consider the light-cone action for Einstein gravity, which can be written in terms of positive and negative helicity fields --- respectively, $h_+$ and $h_-$. To the first interacting order in the gravitational coupling $\kappa$, it takes the form
\begin{align}
S[h_\pm] = \int d^4x\Bigg( & h_- \,\square \,h_+
+ \kappa \; h_-\, \partial_u^2\left( \left(\frac{\partial_w}{\partial_u}h_+\right)^2 -h_+\,\frac{\partial_w^2}{\partial_u^2}h_+ \right) \nonumber \\
& + \kappa \;h_+\, \partial_u^2\left(  \left(\frac{\partial_{\wb}}{\partial_u} h_- \right)^2 -h_-\,\frac{\partial_{\wb}^2}{\partial_u^2} h_- \right) 
 + {\mathcal O}(\kappa^2) \Bigg).
\label{eq:LCgrav}
\end{align}
See e.g.~Appendix~C of \cite{Ananth:2006fh} for the derivation, including the precise origin of $h_\pm$ in terms of metric components. We wrote the action here up to cubic order, but an infinite number of vertices is expected, unlike the case of Yang-Mills theory. What is similar to Yang-Mills theory is that the duality under $h_\pm \mapsto e^{\pm i\theta} h_\pm$ fails at non-linear level. Indeed, the scattering amplitudes violate the selection rule of helicity conservation. This is not surprising given the KLT relations expressing the amplitudes in gravity as a `square' of those in Yang-Mills theory \cite{Kawai:1985xq}. The most compact formula for the MHV amplitudes in gravity was given in \cite{Hodges:2012ym}. A fact about MHV amplitudes related to our discussion is that they can be derived from the amplitude for the helicity flip of a graviton as it crosses a self-dual spacetime background \cite{Mason:2009afn}.

Our conclusion that the duality is broken by interactions agrees with that of ref.~\cite{Deser:2005sz}, which worked up to cubic order in a non-covariant formalism. The procedure employed there is not straightforward, and indeed several authors have since raised the hope that there is an implementation of the U(1) transformation at non-linear level that exhibits the duality. The cubic order is actually subtle in the light-cone action, because the non-local field redefinitions\footnote{Here, $\frac1{\square}$ is understood in terms of the Green's function. This type of field redefinition is related to the notion of MHV Lagrangian in Yang-Mills theory \cite{Mansfield:2005yd,Brandhuber:2006bf}. The gravity story is more intricate \cite{Bianchi:2008pu}.}
\[
\tilde h_\pm = h_\pm
+ \kappa \; \frac1{\square}\,\partial_u^2\left( \left(\frac{\partial_w}{\partial_u}h_\pm\right)^2 -h_\pm\,\frac{\partial_w^2}{\partial_u^2}h_\pm \right)
\]
put the action \eqref{eq:LCgrav} into the form
\[
S[\tilde h_\pm] = \int d^4x\; \left( \tilde h_- \,\square \,\tilde h_+ + {\mathcal O}(\kappa^2) \right)\,.
\]
The duality under $\tilde h_\pm \mapsto e^{\pm i\theta} \tilde h_\pm$ then appears unbroken at this order. This field redefinition is not trivial, and indeed it eliminates the 3-point graviton amplitudes (where $\frac1{\square}$ would be singular); these amplitudes, however, only have on-shell support on complex kinematics. Despite this subtlety at 3 points, the higher-point amplitudes (higher than 4 points) settle the question, due to helicity violation.\footnote{The four-point tree amplitudes preserve helicity. Indeed, `MHV' in this case is two-plus two-minus. When early works appeared applying helicity arguments to gravity amplitudes (e.g.~\cite{Grisaru:1975bx,Grisaru:1976vm}), $(n>4)$-point amplitudes were still too hard to tackle. For this reason, it was often stated that the interactions preserve helicity. A careful reading shows that there is no contradiction.} The breaking of the duality by the interactions is an inescapable conclusion: it is independent of gauge choices and field redefinitions, relying instead on the unambiguous action of the U(1) transformation on the asymptotic scattering states. 

Recently, a proposal was made in \cite{Kol:2022bsd} (see also \cite{Kol:2023yxd}) that the duality holds exactly, which conflicts with our conclusion. The claim there is that there is a way to implement the U(1) transformation that agrees with the linearised story in the appropriate limit, but applies also non-linearly, by mapping any vacuum solution into a vacuum solution. In other words, it is an exact symmetry of the Einstein equations. Consider the curvature 2-form $\mathfrak R_{ab}$ of a vacuum solution. It is claimed in \cite{Kol:2022bsd} that the result of a suitably defined U(1) transformation, $(\mathfrak R_{ab})_\theta$, is also the curvature 2-form of a vacuum solution. This would imply that $(\mathfrak R_{ab})_\theta$ satisfies exactly, for an associated metric, all {\it three} of the crucial relations: equation of motion, first Bianchi identity, and second Bianchi identity. Our argument implies that, for a generic starting solution (we will discuss exceptions), this is not possible. That is, $(\mathfrak R_{ab})_\theta$ is a 2-form, but it cannot be interpreted as the curvature 2-form of a vacuum solution. Consequently, the proposed U(1) transformation is not a map from a generic vacuum solution into another, and it does not imply a U(1) duality symmetry in Einstein gravity.

The first instance we are aware of where the duality in linearised gravity is mentioned, ref.~\cite{Penrose:1960eq}, agrees with our conclusion, while not elaborating on a proof. It is noted there that the Weyl spinor\footnote{Using the spinorial formalism introduced in \cite{Penrose:1960eq}, the self-dual part of the curvature is \,$R^+_{A\dot AB\dot BC\dot CD\dot D}=\frac1{2} \psi_{ABCD}\varepsilon_{\dot A\dot B}\varepsilon_{\dot C\dot D}$\,. Note that, in that paper, $\partial_\mu$ denotes the covariant derivative.} $\psi_{ABCD}$ must satisfy in vacuum the following non-linear wave equation:
\[
\label{eq:psiwave}
\square_g\, \psi_{ABCD} - 3 \, \psi_{(AB}{}^{EF}\, \psi_{CD)EF} =0\,,
\]
where we use $\square_g$ to emphasise the non-trivial metric. The second (differential) Bianchi identity is crucial in the derivation. Unlike the analogous Maxwell case, where the wave equation is linear, the U(1) transformation 
\[
\psi_{ABCD}\,\mapsto\, e^{i\theta}\, \psi_{ABCD}
\]
``does not in general give rise to new exact solutions to the field equations", quoting \cite{Penrose:1960eq}. This statement does not exclude the possibility of a special sector of solutions where the U(1) transformation takes us from one exact vacuum solution to another. It turns out that this may occur even when the second term in \eqref{eq:psiwave} is non-vanishing, because $\square_g$ is also affected by the U(1) transformation. This is the case for the family of solutions to be discussed in the next section.

Before proceeding, let us step back and address an intuitive way in which one may have attempted to define the duality non-linearly in gravity. Suppose that we have initial data ${\mathcal I}$ for which the action of the U(1) transformation is clear, so we have the map ${\mathcal I} \mapsto {\mathcal I}_\theta$. One may say that the solution evolved from the initial data ${\mathcal I}$, on the one hand, and the solution evolved from the initial data ${\mathcal I}_\theta$, on the other hand, are related by a U(1) transformation in some sense, because the two sets of initial data are. But this is not the statement that the theory has the U(1) symmetry. For the theory to have the symmetry, the final data ${\mathcal F}({\mathcal I})$ obtained by evolving ${\mathcal I}$ and the final data ${\mathcal F}({\mathcal I}_\theta)$ obtained by evolving ${\mathcal I}_\theta$ must themselves be mapped under the duality, i.e.~${\mathcal F}({\mathcal I}) \mapsto {\mathcal F}({\mathcal I}_\theta)$. This is precisely the question that we have addressed in terms of scattering amplitudes: we have proven that, generically, ${\mathcal F}({\mathcal I}) \not\mapsto {\mathcal F}({\mathcal I}_\theta)$ for gravity. In the language used in \eqref{eq:SH}, this corresponds to
\[
e^{i\theta{\mathcal H}} S | \text{in}\rangle \not= S e^{i\theta{\mathcal H}}| \text{in}\rangle
\]
for generic $| \text{in}\rangle$ states. That is, ${\mathcal H}$ does not generate a symmetry of the S-matrix. We believe this argument negates the claim in \cite{Compere:2022zdz} that the duality can be realised non-linearly for general vacuum solutions with no incoming radiation. The latter condition means the initial data is specified by canonical multipole moments in perturbation theory, which are then shown to transform naturally under U(1). We claim that, generically, this is merely a transformation of the initial data, and does not represent a symmetry of the dynamics, which would require a time-evolution-compatible U(1) transformation of the final data. This could in principle be tested by the scattering of two Taub-NUT black holes, with no incoming radiation. As we will discuss in the next section, U(1) acts naturally on this initial data. In fact, at leading order, where the scattering is elastic, the duality is obeyed \cite{Huang:2019cja}. The higher perturbative orders of this `nutty' problem are not currently understood, but we expect that the duality fails, because graviton interactions are a crucial ingredient.


\section{A duality-preserving sector of gravity: vacuum type D}
\label{sec:typeD}

We have shown that the duality is broken by the interactions among gravitons. In linearised gravity, which admits the duality as we discussed, the interactions among gravitons are neglected. However, in linearised gravity we still consider interactions with matter, similarly to electromagnetism. We recall that, in electromagnetism, a static dyon with electric charge $e$ and magnetic monopole charge $g$ sources the field strength \eqref{eq:Fdyon}. This is associated to the U(1)-invariant amplitude \eqref{eq:A3EM}. Similarly, in linearised gravity, a static particle with mass $m$ and NUT charge $n$ sources the linearised Taub-NUT solution.\footnote{The notion that this solution is a gravitational dyon goes far back; see e.g.~\cite{Dowker1974TheNS}.} This is associated to the amplitude
\[
\label{eq:A3G}
{\mathcal A}= \frac{m\pm in}{\sqrt{m^2+n^2}}\; \varepsilon_\pm^{\mu\nu} \,p_\mu\, p_\nu
\]
for a `nutty particle' with momentum $p$ to absorb a $\pm$ helicity graviton, $\varepsilon_\pm^{\mu\nu}$ being the latter's polarisation tensor \cite{Huang:2019cja}.\footnote{The momentum of the `nutty particle' is $p=\sqrt{m^2+n^2}\;u$, where $u$ is the velocity.} By considering the duality transformation \eqref{eq:dualityR} of the linearised Riemann tensor for the Taub-NUT solution, we can conclude that 
\[
\label{eq:U1mn}
m- i n \,\mapsto\, e^{ i\theta}(m- i n)= m'- i n'\,.
\]
Together with the duality transformation of the graviton, this makes the amplitude \eqref{eq:A3G} U(1)-invariant, as expected by the duality invariance of the linearised theory.

It turns out that the Taub-NUT family of solutions, labelled by the two parameters $m$ and $n$, admits the duality at the non-linear level. In fact, this property extends straightforwardly to the whole family of vacuum solutions of algebraic type D, labelled by four parameters $(m,n,\gamma,\epsilon)$; the two parameters $\gamma$ and $\epsilon$ combine to represent rotation and acceleration. The explicit expressions are not important for us now, but can be found in \cite{Plebanski:1976gy}. The important point is that, for a natural choice of coordinates, the Weyl spinor takes the form
\[
\label{eq:psitypeD}
\psi_{ABCD} =  (m-in)\, \alpha_{(A}\alpha_B\beta_C\beta_{D)}\,.
\]
Here, the spinors $\alpha_A$ and $\beta_A$ depend on the coordinates, but not on the parameters $(m,n,\gamma,\epsilon)$. While the solution has four parameters, only $m$ and $n$ give rise to curvature. Notice that $\alpha_A\beta^A\neq0$, so the second term in \eqref{eq:psiwave} is present. Nevertheless, from this form of the Weyl spinor, it is clear that the U(1) transformation
\[
\psi_{ABCD}\,\mapsto\, e^{i\theta}\, \psi_{ABCD}
\]
has the effect \eqref{eq:U1mn}, which takes us from the solution with parameters $(m,n,\gamma,\epsilon)$ to the solution with parameters $(m',n',\gamma,\epsilon)$. That is, it is clear in this instance that the U(1) transformation leads to an actual curvature tensor, associated with a metric.\footnote{Had we included a cosmological constant, i.e.~had we considered five parameters $(m,n,\gamma,\epsilon,\Lambda)$, these statements would still hold, except that $\Lambda$ introduces curvature too, via the Ricci tensor as opposed to the Weyl tensor.} As discussed recently in e.g.~\cite{Astorino:2019ljy,Alawadhi:2019urr,Banerjee:2019saj,Astorino:2023elf}, this transformation into a different solution can be interpreted as an Ehlers transformation \cite{Ehlers:1959aug,Geroch:1970nt}.

The argument we gave in the previous section against the non-linear duality is avoided for this family of solutions. The fact that the duality holds here no doubt raised hopes that it would hold more generally in the vacuum solution space of Einstein's theory. These solutions are, however, not generic solutions. They are not algebraically general in the sense of the Petrov classification, but algebraically special --- in particular, of type D, which means they have two principal null directions, associated to $\alpha_A$ and $\beta_A$, each of multiplicity two. A related property is that these solutions admit a double-Kerr-Schild form using complexified coordinates \cite{Plebanski:1976gy}, which means that they have features generically encountered only for linearised solutions. Notice that the form \eqref{eq:psitypeD} of the Weyl spinor is linear in $m$ and $n$. The same special characteristics allow for these exact gravity solutions to be expressed in a very simple manner as a `double copy' of the analogous solutions in electromagnetism \cite{Monteiro:2014cda,Luna:2015paa,Luna:2018dpt}.

We do not know what is the largest class of gravity solutions within which the duality holds. It is clear that it includes also some sector of type N solutions. The results of refs.~\cite{Astorino:2019ljy,Astorino:2023elf} indicate that it may include a family of algebraically general solutions, in fact the full integrable sector of solutions characterised by stationarity and axisymmetry, where the duality is expressed in terms of enhanced Ehlers transformations.
Our expectation is that the solutions in the duality-admitting class are effectively constructible `without interactions among gravitons', which excludes situations of clear physical interest. For instance, interactions among gravitons are crucial in the two-massive-body inspiral or scattering problems, beyond the leading perturbative order.


\section{Conclusion}
\label{sec:conclusion}

We have shown that Einstein gravity does not admit a global U(1) duality symmetry analogous to the electric-magnetic duality in Maxwell theory. This is despite the fact that the symmetry is present (i) in the linearised approximation, and (ii) at non-linear level for a special sector of solutions that notably includes vacuum type D spacetimes, where the duality is known to rotate between mass and NUT charge.

The argument presented here will not be entirely new to some people, but we hope to have fleshed it out for a broad audience, and we are not aware of it being applied to gravity before. The usefulness of scattering amplitudes in gravity, including at the classical level, is now well established; see e.g.~the recent reviews \cite{Bern:2022wqg,Kosower:2022yvp,Bjerrum-Bohr:2022blt,Buonanno:2022pgc,Adamo:2022dcm,DiVecchia:2023frv}, where much work is concerned with gravitational wave physics and/or the double copy. We have given here an example of the power of scattering amplitudes to provide a sharp gauge-invariant answer to a question that has generated some confusion over the years.

Finally, we note that the breaking of the global `electric-magnetic' U(1) symmetry by interactions in gravity is consistent with the lore that a theory of quantum gravity should have no global symmetries; see e.g.~\cite{Banks:1988yz,Banks:2010zn,Harlow:2018tng}. This commonly refers to symmetries of ordinary fields such as the original electric-magnetic duality of Maxwell theory, which is anomalous on a curved spacetime \cite{Agullo:2016lkj}. In our instance of pure Einstein gravity, the global symmetry is broken already at the classical level. It is worth mentioning in this broader context that, in the literature dealing with the ultraviolet behaviour of pure (super)gravities, an explicit connection has been noticed between anomalies and ultraviolet divergences in four dimensions; see e.g.~\cite{Carrasco:2013ypa,Bern:2013uka,Bern:2015xsa,Bern:2017puu,Bern:2017rjw,Freedman:2017zgq,Bern:2019isl,Carrasco:2022lbm}. This and other developments motivate further study of the implications of broken symmetries in gravity.

\section*{Acknowledgements}

The author is grateful to John Joseph Carrasco, George Doran, Uri Kol, Lionel Mason, Nathan Moynihan, Congkao Wen and Sam Wikeley for comments or collaboration on related topics. RM is supported by the Royal Society via a University Research Fellowship.

\bibliography{refs}

\providecommand{\href}[2]{#2}\begingroup\raggedright\begin{thebibliography}{10}

\bibitem{Montonen:1977sn}
C.~Montonen and D.~I. Olive, {\it {Magnetic Monopoles as Gauge Particles?}},
  {\em Phys. Lett. B} {\bf 72} (1977) 117--120.

\bibitem{Seiberg:1994pq}
N.~Seiberg, {\it {Electric - magnetic duality in supersymmetric nonAbelian
  gauge theories}},  {\em Nucl. Phys. B} {\bf 435} (1995) 129--146,
  [\href{http://arxiv.org/abs/hep-th/9411149}{{\tt hep-th/9411149}}].

\bibitem{Gaillard:1981rj}
M.~K. Gaillard and B.~Zumino, {\it {Duality Rotations for Interacting Fields}},
   {\em Nucl. Phys. B} {\bf 193} (1981) 221--244.

\bibitem{Penrose:1960eq}
R.~Penrose, {\it {A Spinor approach to general relativity}},  {\em Annals
  Phys.} {\bf 10} (1960) 171--201.

\bibitem{Nieto:1999pn}
J.~A. Nieto, {\it {S duality for linearized gravity}},  {\em Phys. Lett. A}
  {\bf 262} (1999) 274--281, [\href{http://arxiv.org/abs/hep-th/9910049}{{\tt
  hep-th/9910049}}].

\bibitem{Hull:2001iu}
C.~M. Hull, {\it {Duality in gravity and higher spin gauge fields}},  {\em
  JHEP} {\bf 09} (2001) 027, [\href{http://arxiv.org/abs/hep-th/0107149}{{\tt
  hep-th/0107149}}].

\bibitem{Henneaux:2004jw}
M.~Henneaux and C.~Teitelboim, {\it {Duality in linearized gravity}},  {\em
  Phys. Rev. D} {\bf 71} (2005) 024018,
  [\href{http://arxiv.org/abs/gr-qc/0408101}{{\tt gr-qc/0408101}}].

\bibitem{Deser:2005sz}
S.~Deser and D.~Seminara, {\it {Free spin 2 duality invariance cannot be
  extended to GR}},  {\em Phys. Rev. D} {\bf 71} (2005) 081502,
  [\href{http://arxiv.org/abs/hep-th/0503030}{{\tt hep-th/0503030}}].

\bibitem{Bunster:2006rt}
C.~W. Bunster, S.~Cnockaert, M.~Henneaux, and R.~Portugues, {\it {Monopoles for
  gravitation and for higher spin fields}},  {\em Phys. Rev. D} {\bf 73} (2006)
  105014, [\href{http://arxiv.org/abs/hep-th/0601222}{{\tt hep-th/0601222}}].

\bibitem{Argurio:2009xr}
R.~Argurio and F.~Dehouck, {\it {Gravitational duality and rotating
  solutions}},  {\em Phys. Rev. D} {\bf 81} (2010) 064010,
  [\href{http://arxiv.org/abs/0909.0542}{{\tt arXiv:0909.0542}}].

\bibitem{Barnich:2008ts}
G.~Barnich and C.~Troessaert, {\it {Manifest spin 2 duality with electric and
  magnetic sources}},  {\em JHEP} {\bf 01} (2009) 030,
  [\href{http://arxiv.org/abs/0812.0552}{{\tt arXiv:0812.0552}}].

\bibitem{deHaro:2008gp}
S.~de~Haro, {\it {Dual Gravitons in AdS(4) / CFT(3) and the Holographic Cotton
  Tensor}},  {\em JHEP} {\bf 01} (2009) 042,
  [\href{http://arxiv.org/abs/0808.2054}{{\tt arXiv:0808.2054}}].

\bibitem{Astorino:2019ljy}
M.~Astorino, {\it {Enhanced Ehlers Transformation and the
  Majumdar-Papapetrou-NUT Spacetime}},  {\em JHEP} {\bf 01} (2020) 123,
  [\href{http://arxiv.org/abs/1906.08228}{{\tt arXiv:1906.08228}}].

\bibitem{Boos:2021suz}
J.~Boos and I.~Kol\'a\v{r}, {\it {Nonlocality and gravitoelectromagnetic
  duality}},  {\em Phys. Rev. D} {\bf 104} (2021), no.~2 024018,
  [\href{http://arxiv.org/abs/2103.10555}{{\tt arXiv:2103.10555}}].

\bibitem{Astorino:2023elf}
M.~Astorino and G.~Boldi, {\it {Plebanski-Demianski goes NUTs (to remove the
  Misner string)}},  {\em JHEP} {\bf 08} (2023) 085,
  [\href{http://arxiv.org/abs/2305.03744}{{\tt arXiv:2305.03744}}].

\bibitem{Deser:1976iy}
S.~Deser and C.~Teitelboim, {\it {Duality Transformations of Abelian and
  Nonabelian Gauge Fields}},  {\em Phys. Rev. D} {\bf 13} (1976) 1592--1597.

\bibitem{Kol:2022bsd}
U.~Kol, {\it {Duality in Einstein's Gravity}},
  \href{http://arxiv.org/abs/2205.05752}{{\tt arXiv:2205.05752}}.

\bibitem{Kol:2023yxd}
U.~Kol and S.-T. Yau, {\it {Duality in Gauge Theory, Gravity and String
  Theory}},  \href{http://arxiv.org/abs/2311.07934}{{\tt arXiv:2311.07934}}.

\bibitem{Luna:2015paa}
A.~Luna, R.~Monteiro, D.~O'Connell, and C.~D. White, {\it {The classical double
  copy for Taub-NUT spacetime}},  {\em Phys. Lett.} {\bf B750} (2015) 272--277,
  [\href{http://arxiv.org/abs/1507.01869}{{\tt arXiv:1507.01869}}].

\bibitem{Huang:2019cja}
Y.-T. Huang, U.~Kol, and D.~O'Connell, {\it {Double copy of electric-magnetic
  duality}},  {\em Phys. Rev. D} {\bf 102} (2020), no.~4 046005,
  [\href{http://arxiv.org/abs/1911.06318}{{\tt arXiv:1911.06318}}].

\bibitem{Alawadhi:2019urr}
R.~Alawadhi, D.~S. Berman, B.~Spence, and D.~Peinador~Veiga, {\it {S-duality
  and the double copy}},  {\em JHEP} {\bf 03} (2020) 059,
  [\href{http://arxiv.org/abs/1911.06797}{{\tt arXiv:1911.06797}}].

\bibitem{Banerjee:2019saj}
A.~Banerjee, E.~O. Colg\'ain, J.~A. Rosabal, and H.~Yavartanoo, {\it {Ehlers as
  EM duality in the double copy}},  {\em Phys. Rev. D} {\bf 102} (2020) 126017,
  [\href{http://arxiv.org/abs/1912.02597}{{\tt arXiv:1912.02597}}].

\bibitem{Emond:2020lwi}
W.~T. Emond, Y.-T. Huang, U.~Kol, N.~Moynihan, and D.~O'Connell, {\it
  {Amplitudes from Coulomb to Kerr-Taub-NUT}},  {\em JHEP} {\bf 05} (2022) 055,
  [\href{http://arxiv.org/abs/2010.07861}{{\tt arXiv:2010.07861}}].

\bibitem{Moynihan:2020gxj}
N.~Moynihan and J.~Murugan, {\it {On-Shell Electric-Magnetic Duality and the
  Dual Graviton}},  \href{http://arxiv.org/abs/2002.11085}{{\tt
  arXiv:2002.11085}}.

\bibitem{Monteiro:2020plf}
R.~Monteiro, D.~O'Connell, D.~Peinador~Veiga, and M.~Sergola, {\it {Classical
  solutions and their double copy in split signature}},  {\em JHEP} {\bf 05}
  (2021) 268, [\href{http://arxiv.org/abs/2012.11190}{{\tt arXiv:2012.11190}}].

\bibitem{Crawley:2021auj}
E.~Crawley, A.~Guevara, N.~Miller, and A.~Strominger, {\it {Black holes in
  Klein space}},  {\em JHEP} {\bf 10} (2022) 135,
  [\href{http://arxiv.org/abs/2112.03954}{{\tt arXiv:2112.03954}}].

\bibitem{Monteiro:2021ztt}
R.~Monteiro, S.~Nagy, D.~O'Connell, D.~Peinador~Veiga, and M.~Sergola, {\it
  {NS-NS spacetimes from amplitudes}},  {\em JHEP} {\bf 06} (2022) 021,
  [\href{http://arxiv.org/abs/2112.08336}{{\tt arXiv:2112.08336}}].

\bibitem{Compere:2022zdz}
G.~Comp\`ere, R.~Oliveri, and A.~Seraj, {\it {Metric reconstruction from
  celestial multipoles}},  {\em JHEP} {\bf 11} (2022) 001,
  [\href{http://arxiv.org/abs/2206.12597}{{\tt arXiv:2206.12597}}].

\bibitem{Calkin:1965}
M.~G. Calkin, {\it An invariance property of the free electromagnetic field},
  {\em Am. J. Phys.} {\bf 33} (1965) 958--960.

\bibitem{Novotny:2018iph}
J.~Novotn\'y, {\it {Self-duality, helicity conservation and normal ordering in
  nonlinear QED}},  {\em Phys. Rev. D} {\bf 98} (2018), no.~8 085015,
  [\href{http://arxiv.org/abs/1806.02167}{{\tt arXiv:1806.02167}}].

\bibitem{Rosly:2002jt}
A.~A. Rosly and K.~G. Selivanov, {\it {Helicity conservation in Born-Infeld
  theory}},  in {\em {Workshop on String Theory and Complex Geometry}}, 4,
  2002.
\newblock \href{http://arxiv.org/abs/hep-th/0204229}{{\tt hep-th/0204229}}.

\bibitem{Carrasco:2013ypa}
J.~Carrasco, R.~Kallosh, R.~Roiban, and A.~Tseytlin, {\it {On the U(1) duality
  anomaly and the S-matrix of N=4 supergravity}},  {\em JHEP} {\bf 1307} (2013)
  029, [\href{http://arxiv.org/abs/1303.6219}{{\tt arXiv:1303.6219}}].

\bibitem{Bern:2013uka}
Z.~Bern, S.~Davies, T.~Dennen, A.~V. Smirnov, and V.~A. Smirnov, {\it
  {Ultraviolet Properties of N=4 Supergravity at Four Loops}},  {\em
  Phys.Rev.Lett.} {\bf 111} (2013), no.~23 231302,
  [\href{http://arxiv.org/abs/1309.2498}{{\tt arXiv:1309.2498}}].

\bibitem{Freedman:2017zgq}
D.~Z. Freedman, R.~Kallosh, D.~Murli, A.~Van~Proeyen, and Y.~Yamada, {\it
  {Absence of U(1) Anomalous Superamplitudes in $\mathcal{N}\geq 5$
  Supergravities}},  {\em JHEP} {\bf 05} (2017) 067,
  [\href{http://arxiv.org/abs/1703.03879}{{\tt arXiv:1703.03879}}].

\bibitem{Bern:2017rjw}
Z.~Bern, J.~Parra-Martinez, and R.~Roiban, {\it {Canceling the U(1) Anomaly in
  the $S$ Matrix of $N$=4 Supergravity}},  {\em Phys. Rev. Lett.} {\bf 121}
  (2018), no.~10 101604, [\href{http://arxiv.org/abs/1712.03928}{{\tt
  arXiv:1712.03928}}].

\bibitem{Pavao:2022kog}
N.~H. Pavao, {\it {Effective observables for electromagnetic duality from novel
  amplitude decomposition}},  {\em Phys. Rev. D} {\bf 107} (2023), no.~6
  065020, [\href{http://arxiv.org/abs/2210.12800}{{\tt arXiv:2210.12800}}].

\bibitem{Carrasco:2022jxn}
J.~J.~M. Carrasco and N.~H. Pavao, {\it {Virtues of a symmetric-structure
  double copy}},  {\em Phys. Rev. D} {\bf 107} (2023), no.~6 065005,
  [\href{http://arxiv.org/abs/2211.04431}{{\tt arXiv:2211.04431}}].

\bibitem{Carrasco:2023qgz}
J.~J.~M. Carrasco and N.~H. Pavao, {\it {Even-point Multi-loop Unitarity and
  its Applications: Exponentiation, Anomalies and Evanescence}},
  \href{http://arxiv.org/abs/2307.16812}{{\tt arXiv:2307.16812}}.

\bibitem{Agullo:2016lkj}
I.~Agullo, A.~del Rio, and J.~Navarro-Salas, {\it {Electromagnetic duality
  anomaly in curved spacetimes}},  {\em Phys. Rev. Lett.} {\bf 118} (2017),
  no.~11 111301, [\href{http://arxiv.org/abs/1607.08879}{{\tt
  arXiv:1607.08879}}].

\bibitem{Bengtsson:1983pg}
A.~K.~H. Bengtsson, I.~Bengtsson, and L.~Brink, {\it {Cubic Interaction Terms
  for Arbitrarily Extended Supermultiplets}},  {\em Nucl. Phys. B} {\bf 227}
  (1983) 41--49.

\bibitem{Gibbons:1995cv}
G.~W. Gibbons and D.~A. Rasheed, {\it {Electric - magnetic duality rotations in
  nonlinear electrodynamics}},  {\em Nucl. Phys. B} {\bf 454} (1995) 185--206,
  [\href{http://arxiv.org/abs/hep-th/9506035}{{\tt hep-th/9506035}}].

\bibitem{Elvang:2019twd}
H.~Elvang, M.~Hadjiantonis, C.~R.~T. Jones, and S.~Paranjape, {\it
  {All-Multiplicity One-Loop Amplitudes in Born-Infeld Electrodynamics from
  Generalized Unitarity}},  {\em JHEP} {\bf 03} (2020) 009,
  [\href{http://arxiv.org/abs/1906.05321}{{\tt arXiv:1906.05321}}].

\bibitem{Wen:2020qrj}
C.~Wen and S.-Q. Zhang, {\it {D3-Brane Loop Amplitudes from M5-Brane Tree
  Amplitudes}},  {\em JHEP} {\bf 07} (2020) 098,
  [\href{http://arxiv.org/abs/2004.02735}{{\tt arXiv:2004.02735}}].

\bibitem{Elvang:2020kuj}
H.~Elvang, M.~Hadjiantonis, C.~R.~T. Jones, and S.~Paranjape, {\it
  {Electromagnetic Duality and D3-Brane Scattering Amplitudes Beyond Leading
  Order}},  {\em JHEP} {\bf 04} (2021) 173,
  [\href{http://arxiv.org/abs/2006.08928}{{\tt arXiv:2006.08928}}].

\bibitem{Cachazo:2014xea}
F.~Cachazo, S.~He, and E.~Y. Yuan, {\it {Scattering Equations and Matrices:
  From Einstein To Yang-Mills, DBI and NLSM}},  {\em JHEP} {\bf 07} (2015) 149,
  [\href{http://arxiv.org/abs/1412.3479}{{\tt arXiv:1412.3479}}].

\bibitem{Bern:1999bx}
Z.~Bern, A.~De~Freitas, and H.~L. Wong, {\it {On the coupling of gravitons to
  matter}},  {\em Phys. Rev. Lett.} {\bf 84} (2000) 3531,
  [\href{http://arxiv.org/abs/hep-th/9912033}{{\tt hep-th/9912033}}].

\bibitem{Chiodaroli:2014xia}
M.~Chiodaroli, M.~Gunaydin, H.~Johansson, and R.~Roiban, {\it {Scattering
  amplitudes in N=2 Maxwell-Einstein and Yang-Mills/Einstein supergravity}},
  \href{http://arxiv.org/abs/1408.0764}{{\tt arXiv:1408.0764}}.

\bibitem{Gibbons:1995ap}
G.~W. Gibbons and D.~A. Rasheed, {\it {Sl(2,R) invariance of nonlinear
  electrodynamics coupled to an axion and a dilaton}},  {\em Phys. Lett. B}
  {\bf 365} (1996) 46--50, [\href{http://arxiv.org/abs/hep-th/9509141}{{\tt
  hep-th/9509141}}].

\bibitem{Cano:2021tfs}
P.~A. Cano and A.~Murcia, {\it {Duality-invariant extensions of
  Einstein-Maxwell theory}},  {\em JHEP} {\bf 08} (2021) 042,
  [\href{http://arxiv.org/abs/2104.07674}{{\tt arXiv:2104.07674}}].

\bibitem{Britto:2005fq}
R.~Britto, F.~Cachazo, B.~Feng, and E.~Witten, {\it {Direct proof of tree-level
  recursion relation in Yang-Mills theory}},  {\em Phys. Rev. Lett.} {\bf 94}
  (2005) 181602, [\href{http://arxiv.org/abs/hep-th/0501052}{{\tt
  hep-th/0501052}}].

\bibitem{Arkani-Hamed:2012zlh}
N.~Arkani-Hamed, J.~L. Bourjaily, F.~Cachazo, A.~B. Goncharov, A.~Postnikov,
  and J.~Trnka, {\em {Grassmannian Geometry of Scattering Amplitudes}}.
\newblock Cambridge University Press, 4, 2016.

\bibitem{Bern:2011qt}
Z.~Bern and Y.-t. Huang, {\it {Basics of Generalized Unitarity}},  {\em
  J.Phys.A} {\bf A44} (2011) 454003,
  [\href{http://arxiv.org/abs/1103.1869}{{\tt arXiv:1103.1869}}].

\bibitem{Csaki:2020inw}
C.~Csaki, S.~Hong, Y.~Shirman, O.~Telem, J.~Terning, and M.~Waterbury, {\it
  {Scattering amplitudes for monopoles: pairwise little group and pairwise
  helicity}},  {\em JHEP} {\bf 08} (2021) 029,
  [\href{http://arxiv.org/abs/2009.14213}{{\tt arXiv:2009.14213}}].

\bibitem{Kosower:2018adc}
D.~A. Kosower, B.~Maybee, and D.~O'Connell, {\it {Amplitudes, Observables, and
  Classical Scattering}},  {\em JHEP} {\bf 02} (2019) 137,
  [\href{http://arxiv.org/abs/1811.10950}{{\tt arXiv:1811.10950}}].

\bibitem{Parke:1986gb}
S.~J. Parke and T.~R. Taylor, {\it {An Amplitude for $n$ Gluon Scattering}},
  {\em Phys. Rev. Lett.} {\bf 56} (1986) 2459.

\bibitem{Chalmers:1998jb}
G.~Chalmers and W.~Siegel, {\it {Simplifying algebra in Feynman graphs. Part 2.
  Spinor helicity from the space-cone}},  {\em Phys. Rev. D} {\bf 59} (1999)
  045013, [\href{http://arxiv.org/abs/hep-ph/9801220}{{\tt hep-ph/9801220}}].

\bibitem{Ananth:2006fh}
S.~Ananth, L.~Brink, R.~Heise, and H.~G. Svendsen, {\it {The N=8 Supergravity
  Hamiltonian as a Quadratic Form}},  {\em Nucl. Phys. B} {\bf 753} (2006)
  195--210, [\href{http://arxiv.org/abs/hep-th/0607019}{{\tt hep-th/0607019}}].

\bibitem{Kawai:1985xq}
H.~Kawai, D.~Lewellen, and S.~Tye, {\it {A Relation Between Tree Amplitudes of
  Closed and Open Strings}},  {\em Nucl.Phys.} {\bf B269} (1986) 1.

\bibitem{Hodges:2012ym}
A.~Hodges, {\it {A simple formula for gravitational MHV amplitudes}},
  \href{http://arxiv.org/abs/1204.1930}{{\tt arXiv:1204.1930}}.

\bibitem{Mason:2009afn}
L.~J. Mason and D.~Skinner, {\it {Gravity, Twistors and the MHV Formalism}},
  {\em Commun. Math. Phys.} {\bf 294} (2010) 827--862,
  [\href{http://arxiv.org/abs/0808.3907}{{\tt arXiv:0808.3907}}].

\bibitem{Mansfield:2005yd}
P.~Mansfield, {\it {The Lagrangian origin of MHV rules}},  {\em JHEP} {\bf 03}
  (2006) 037, [\href{http://arxiv.org/abs/hep-th/0511264}{{\tt
  hep-th/0511264}}].

\bibitem{Brandhuber:2006bf}
A.~Brandhuber, B.~Spence, and G.~Travaglini, {\it {Amplitudes in Pure
  Yang-Mills and MHV Diagrams}},  {\em JHEP} {\bf 02} (2007) 088,
  [\href{http://arxiv.org/abs/hep-th/0612007}{{\tt hep-th/0612007}}].

\bibitem{Bianchi:2008pu}
M.~Bianchi, H.~Elvang, and D.~Z. Freedman, {\it {Generating Tree Amplitudes in
  N=4 SYM and N = 8 SG}},  {\em JHEP} {\bf 09} (2008) 063,
  [\href{http://arxiv.org/abs/0805.0757}{{\tt arXiv:0805.0757}}].

\bibitem{Grisaru:1975bx}
M.~T. Grisaru, P.~van Nieuwenhuizen, and C.~C. Wu, {\it {Gravitational Born
  Amplitudes and Kinematical Constraints}},  {\em Phys. Rev. D} {\bf 12} (1975)
  397.

\bibitem{Grisaru:1976vm}
M.~T. Grisaru, H.~N. Pendleton, and P.~van Nieuwenhuizen, {\it {Supergravity
  and the S Matrix}},  {\em Phys. Rev. D} {\bf 15} (1977) 996.

\bibitem{Dowker1974TheNS}
J.~S. Dowker, {\it The nut solution as a gravitational dyon},  {\em General
  Relativity and Gravitation} {\bf 5} (1974) 603--613.

\bibitem{Plebanski:1976gy}
J.~F. Plebanski and M.~Demianski, {\it {Rotating, charged, and uniformly
  accelerating mass in general relativity}},  {\em Annals Phys.} {\bf 98}
  (1976) 98--127.

\bibitem{Ehlers:1959aug}
J.~Ehlers, {\it {Transformations of static exterior solutions of Einstein's
  gravitational field equations into different solutions by means of conformal
  mapping}},  {\em Colloq. Int. CNRS} {\bf 91} (1962) 275--284.

\bibitem{Geroch:1970nt}
R.~P. Geroch, {\it {A Method for generating solutions of Einstein's
  equations}},  {\em J. Math. Phys.} {\bf 12} (1971) 918--924.

\bibitem{Monteiro:2014cda}
R.~Monteiro, D.~O'Connell, and C.~D. White, {\it {Black holes and the double
  copy}},  {\em JHEP} {\bf 1412} (2014) 056,
  [\href{http://arxiv.org/abs/1410.0239}{{\tt arXiv:1410.0239}}].

\bibitem{Luna:2018dpt}
A.~Luna, R.~Monteiro, I.~Nicholson, and D.~O'Connell, {\it {Type D Spacetimes
  and the Weyl Double Copy}},  {\em Class. Quant. Grav.} {\bf 36} (2019)
  065003, [\href{http://arxiv.org/abs/1810.08183}{{\tt arXiv:1810.08183}}].

\bibitem{Bern:2022wqg}
Z.~Bern, J.~J. Carrasco, M.~Chiodaroli, H.~Johansson, and R.~Roiban, {\it {The
  SAGEX Review on Scattering Amplitudes, Chapter 2: An Invitation to
  Color-Kinematics Duality and the Double Copy}},
  \href{http://arxiv.org/abs/2203.13013}{{\tt arXiv:2203.13013}}.

\bibitem{Kosower:2022yvp}
D.~A. Kosower, R.~Monteiro, and D.~O'Connell, {\it {The SAGEX review on
  scattering amplitudes Chapter 14: Classical gravity from scattering
  amplitudes}},  {\em J. Phys. A} {\bf 55} (2022), no.~44 443015,
  [\href{http://arxiv.org/abs/2203.13025}{{\tt arXiv:2203.13025}}].

\bibitem{Bjerrum-Bohr:2022blt}
N.~E.~J. Bjerrum-Bohr, P.~H. Damgaard, L.~Plante, and P.~Vanhove, {\it {The
  SAGEX review on scattering amplitudes Chapter 13: Post-Minkowskian expansion
  from scattering amplitudes}},  {\em J. Phys. A} {\bf 55} (2022), no.~44
  443014, [\href{http://arxiv.org/abs/2203.13024}{{\tt arXiv:2203.13024}}].

\bibitem{Buonanno:2022pgc}
A.~Buonanno, M.~Khalil, D.~O'Connell, R.~Roiban, M.~P. Solon, and M.~Zeng, {\it
  {Snowmass White Paper: Gravitational Waves and Scattering Amplitudes}},  in
  {\em {Snowmass 2021}}, 4, 2022.
\newblock \href{http://arxiv.org/abs/2204.05194}{{\tt arXiv:2204.05194}}.

\bibitem{Adamo:2022dcm}
T.~Adamo, J.~J.~M. Carrasco, M.~Carrillo-Gonz\'alez, M.~Chiodaroli, H.~Elvang,
  H.~Johansson, D.~O'Connell, R.~Roiban, and O.~Schlotterer, {\it {Snowmass
  White Paper: the Double Copy and its Applications}},  in {\em {2022 Snowmass
  Summer Study}}, 4, 2022.
\newblock \href{http://arxiv.org/abs/2204.06547}{{\tt arXiv:2204.06547}}.

\bibitem{DiVecchia:2023frv}
P.~Di~Vecchia, C.~Heissenberg, R.~Russo, and G.~Veneziano, {\it {The
  gravitational eikonal: from particle, string and brane collisions to
  black-hole encounters}},  \href{http://arxiv.org/abs/2306.16488}{{\tt
  arXiv:2306.16488}}.

\bibitem{Banks:1988yz}
T.~Banks and L.~J. Dixon, {\it {Constraints on String Vacua with Space-Time
  Supersymmetry}},  {\em Nucl. Phys. B} {\bf 307} (1988) 93--108.

\bibitem{Banks:2010zn}
T.~Banks and N.~Seiberg, {\it {Symmetries and Strings in Field Theory and
  Gravity}},  {\em Phys. Rev. D} {\bf 83} (2011) 084019,
  [\href{http://arxiv.org/abs/1011.5120}{{\tt arXiv:1011.5120}}].

\bibitem{Harlow:2018tng}
D.~Harlow and H.~Ooguri, {\it {Symmetries in quantum field theory and quantum
  gravity}},  {\em Commun. Math. Phys.} {\bf 383} (2021), no.~3 1669--1804,
  [\href{http://arxiv.org/abs/1810.05338}{{\tt arXiv:1810.05338}}].

\bibitem{Bern:2015xsa}
Z.~Bern, C.~Cheung, H.-H. Chi, S.~Davies, L.~Dixon, and J.~Nohle, {\it
  {Evanescent Effects Can Alter Ultraviolet Divergences in Quantum Gravity
  without Physical Consequences}},  {\em Phys. Rev. Lett.} {\bf 115} (2015),
  no.~21 211301, [\href{http://arxiv.org/abs/1507.06118}{{\tt
  arXiv:1507.06118}}].

\bibitem{Bern:2017puu}
Z.~Bern, H.-H. Chi, L.~Dixon, and A.~Edison, {\it {Two-Loop Renormalization of
  Quantum Gravity Simplified}},  {\em Phys. Rev. D} {\bf 95} (2017), no.~4
  046013, [\href{http://arxiv.org/abs/1701.02422}{{\tt arXiv:1701.02422}}].

\bibitem{Bern:2019isl}
Z.~Bern, D.~Kosower, and J.~Parra-Martinez, {\it {Two-loop n-point anomalous
  amplitudes in N=4 supergravity}},  {\em Proc. Roy. Soc. Lond. A} {\bf 476}
  (2020), no.~2235 20190722, [\href{http://arxiv.org/abs/1905.05151}{{\tt
  arXiv:1905.05151}}].

\bibitem{Carrasco:2022lbm}
J.~J.~M. Carrasco, M.~Lewandowski, and N.~H. Pavao, {\it {The color-dual fate
  of N=4 supergravity}},  \href{http://arxiv.org/abs/2203.03592}{{\tt
  arXiv:2203.03592}}.

\end{thebibliography}\endgroup
\bibliographystyle{JHEP}

\end{document}